\documentclass[twocolumn,aps,showpacs,superscriptaddress]{revtex4}
\usepackage{graphicx}% Include Fig. files
\usepackage{dcolumn}% Align table columns on decimal point
\usepackage{bm}% bold math
\usepackage{xspace}
\usepackage{multirow}
\usepackage{natbib}
\usepackage{color}

\renewcommand{\vec}{\mathbf}

\newcommand{\bioned}{Bi$_{2}$Sr$_2$CuO$_{6+\delta}$}

\begin{document}

\preprint{}

\title{Evolution of Fermi surface and normal-state gap in chemically
substituted cuprates Bi$_{2}$Sr$_{2-x}$Bi$_{x}$CuO$_{6+\delta}$}

\author{Z.-H. Pan}
\affiliation{Department of Physics, Boston College, Chestnut Hill, MA 02467}
\author{P. Richard}
\affiliation{Department of Physics, Boston College, Chestnut Hill, MA 02467}
\author{Y.-M. Xu}
\affiliation{Department of Physics, Boston College, Chestnut Hill, MA 02467}
\author{M. Neupane}
\affiliation{Department of Physics, Boston College, Chestnut Hill, MA 02467}
\author{P. Bishay}
\affiliation{Department of Physics, Boston College, Chestnut Hill, MA 02467}
\author{A. V. Fedorov}
\affiliation{Advanced Light Source, Lawrence Berkeley National Laboratory, Berkeley, CA 94720}
\author{H. Luo}
\affiliation{National Laboratory for Superconductivity, Institute of Physics and National Laboratory for Condensed Matter Physics, P. O. Box 603 Beijing, 100080, P. R. China}
\author{L. Fang}
\affiliation{National Laboratory for Superconductivity, Institute of Physics and National Laboratory for Condensed Matter Physics, P. O. Box 603 Beijing, 100080, P. R. China}
\author{H.-H. Wen}
\affiliation{National Laboratory for Superconductivity, Institute of Physics and National Laboratory for Condensed Matter Physics, P. O. Box 603 Beijing, 100080, P. R. China}
\author{Z. Wang}
\affiliation{Department of Physics, Boston College, Chestnut Hill, MA 02467}
\author{H. Ding}\email{dingh@bc.edu}
\affiliation{Department of Physics, Boston College, Chestnut Hill, MA 02467}

\date{\today}% It is always \today, today,

\begin{abstract}
We have performed a systematic angle-resolved photoemission study of
chemically substituted cuprates
Bi$_{2}$Sr$_{2-x}$Bi$_{x}$CuO$_{6+\delta}$. We observed that the Fermi
surface area shrinks linearly with Bi substitution content $x$,
reflecting the electron doping nature of this chemical substitution.
In addition, the spectral linewidth broadens rapidly with increasing
$x$, and becomes completely incoherent at the
superconducting-insulating boundary. The $d$-wave-like normal-state
gap observed in the lightly underdoped region gradually evolves into
a large soft gap, which suppresses antinodal spectral weight
linearly in both the excitation energy and temperature. Combining
with the bulk resistivity data obtained on the same samples, we
establish the emergence of the Coulomb gap behavior in the very
underdoped regime. Our results reveal the dual roles, doping and
disorder, of off-plane chemical substitutions in high-$T_c$ cuprates
and elucidate the nature of the quantum electronic states due to
strong correlation and disorder.
\end{abstract}
\vspace{1.0cm}


\maketitle
\pagebreak

Since high temperature superconductivity in cuprates is achieved by
adding extra carriers into an antiferromagnetic Mott insulator, how
the electronic structures, such as Fermi surface (FS) and energy gap
(including the superconducting gap and the normal state pseudogap),
evolve with doping is critical to understanding the elusive
superconducting mechanism. Direct spectroscopic probe of the
electronic structures, such as angle-resolved photoemission
spectroscopy (ARPES) and scanning tunneling microscopy (STM), are
mostly done on the Bi$_2$Sr$_2$Ca$_n$Cu$_{n+1}$O$_{2n+6}$ compounds,
whose doping range is limited through oxygenation. More heavily
underdoping can only be achieved by chemical substitution
\cite{Tanakat,Kanigel}. By substituting trivalent Bi ions for
divalent Sr ions into the apical plane of \bioned (Bi2201), we have
further pushed the underdoping threshold continuously to the
insulating phase \cite{Wen}. We establish a concrete linear
relationship between the reduction of the hole concentration and the
Bi substitution through ARPES by measuring the reduction of the
Luttinger volume enclosed by the FS. In addition, we
observe that quasiparticle coherence vanishes at the
superconducting-insulating boundary. More importantly, the
$d$-wave-like pseudogap observed in the lightly underdoped region
gradually evolves into a large soft gap, which suppresses antinodal
spectral weight linearly in both the excitation energy and
temperature. We propose that the pseudogap in the underdoped regime
evolves into a Coulomb gap beyond the superconducting-insulating
phase boundary. To support this finding, we performed transport
measurements on the same samples and found that the resistivity data
are consistent with hopping transport in the presence of a Coulomb
gap in the highly underdoped regime, characteristic of strongly
disordered correlated systems. The Coulomb energy scale derived from
transport agrees remarkably well with the gap energy in ARPES
spectra. The observed universality of this pseudogap evolution
reveals the dual roles, underdoping and disorder, of chemical
substitution in high-$T_c$ cuprates.

High quality single-crystals of
Bi$_{2}$Sr$_{2-x}$Bi$_{x}$CuO$_{6+\delta}$ (Bi-Bi2201) were grown by the
traveling solvent floating zone method. $T_c$ of superconducting
samples were determined by AC susceptibility, the superconducting
transition width is about 0.5 - 1.3 K. More technical details have
been described previously \cite{Wen}. ARPES experiments were
performed at the beamline U1NIM of the Synchrotron Radiation Center
in Wisconsin, and the beamline 12.0.1 of the Advanced Light Source in
California. Energy and momentum resolutions have been set to $\sim$
10 - 20 meV and $\sim0.02$ \AA$^{-1}$, respectively. All samples
were cleaved and measured \emph{in situ} in a vacuum better than
$8\times10^{-11}$ Torr on a flat (001) surface.

\begin{figure}[htbp]
\begin{center}
\includegraphics[width=8cm]{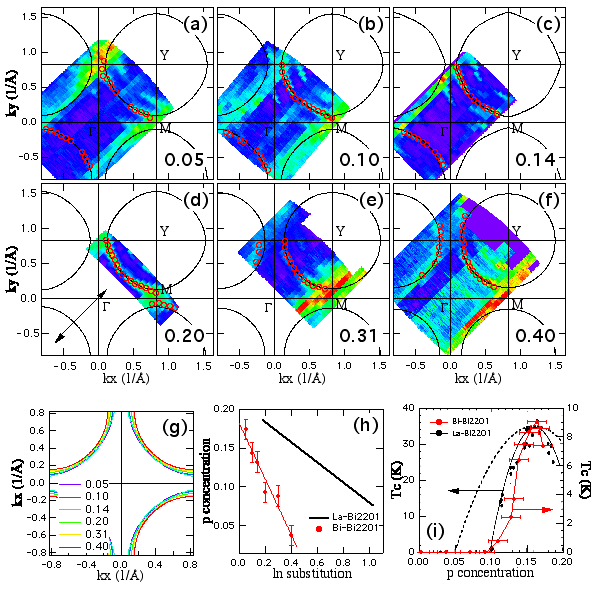}
\caption{\label{mapping} FS mapping of Bi$_{2}$Sr$_{2-x}$Bi$_x$CuO$_{6+\delta}$ at different Bi
contents. (a)-(f) Plots of ARPES intensity integrated within
$\omega=0\pm 12$ meV for $x$ = 0.05, 0.1, 0.14, 0.2, 0.31, 0.4. The
red circles are extracted from MDC peaks, and the lines are the
tight-binding fits to the circles. All the samples are aligned along
$\Gamma$Y parallel to $\vec{A}$, as shown by the arrow in (d). (g)
FS contours from tight-binding fits. (h) Doping effect induced by
the substitution of Bi (dots, the thin line is the best linear fit)
and La (solid line) measured by Hall measurements \cite{Ando}. (i)
Corresponding phase diagrams of
Bi-Bi2201 and
La-Bi2201, in comparison to a generic
phase diagram of the high-$T_c$ cuprates \cite{phasediagram}.}
\end{center}
\end{figure}

We show in Fig.~\ref{mapping} the evolution of the Fermi surface
of Bi$_{2}$Sr$_{2-x}$Bi$_x$CuO$_{6+\delta}$ at 20 K
for six different Bi contents from $x$ = 0.05 to 0.4. The Fermi
vectors extracted from the momentum distribution curve (MDC) peaks
are consistent with the intensity plots and allow a precise
determination of the underlying FS, which is defined as a contour of
the minimum gap locus \cite{MGL}. We report in Fig.~ \ref{mapping}g
all the underlying FS extracted for the various Bi concentrations
using an effective tight-binding fit. We notice a smooth evolution
in the shape of the FS. In particular, the FS at the antinodes moves
away from the M($\pi$, 0) point as the Bi content increases. More
importantly, as shown more clearly in Fig.~\ref{mapping}g, the size
of the Y($\pi$,$\pi$)-centered holelike FS decreases continuously
with the Bi concentration, indicating the underdoping nature of the
Sr$^{2+}$$\rightarrow$Bi$^{3+}$ substitution.

%Fig 1 #######################################################################
% #######################################################################

We extract experimental values of the effective hole-doping $p$ and
plot them as a function of the Bi substitution $x$ in
Fig.~\ref{mapping}h, by using Luttinger theorem, which states that
the volume of an enclosed FS is proportional to the carrier
concentration 1+$p$. While the derived doping $p$ decreases linearly
with $x$, as shown in Fig.~\ref{mapping}h, the linear fit gives
$p=0.182-0.36x$, which indicates that each substituting Bi atom
removes only 0.36 holelike carrier from the CuO$_2$ plane. A similar
result, albeit with an even smaller doping efficiency, has been
observed in Bi$_{2}$Sr$_{2-x}$La$_x$CuO$_{6+\delta}$ (La-Bi2201) by
Hall measurements \cite{Ando}, also plotted in Fig.~\ref{mapping}h
as comparison. The reason for this surprising behavior is not clear
yet. Although one would expect that instead of the usual trivalent
ions, the divalent ions of Bi or La could be present, measurements
of Bi core levels in our samples are inconsistent with this
hypothesis. One possibility to explain this phenomenon is that the
oxygen dopant concentration $\delta$, which is not known precisely
in these materials, increases with $x$, compensating the extra
charge of the Bi$^{3+}$ ion, and maintaining total charge neutrality
$2\delta = p+x$. Spectroscopic evidence has led to a similar
scenario for the electron-doped cuprates, where the Ce$^{4+}$ dopant
ions tend to form pairs with the extraneous oxygen ions \cite{Riou}.

In Fig.~\ref{mapping}i, we construct the phase diagram of
Bi-Bi2201 based on
transport/susceptibility measurements \cite{Wen} and the
relationship of $p$ vs $x$ displayed in Fig.~\ref{mapping}h. This
phase diagram, when rescaled along the temperature axis, matches
well with the one constructed from the Hall measurements of the
La-Bi2201 \cite{Ando}. We note that in both Bi and La substituted
samples, $T_c$ vanishes around 10\% hole concentration, which is
larger than the 5\% critical value found in many cuprates
\cite{phasediagram}. This suggests that underdoping is not the only
effect introduced by the Sr$^{2+}\rightarrow$Bi$^{3+}$ substitution.
Beyond the chemical doping necessary to vary the carrier
concentration in cuprates, a host of experiments have provided
evidences that the ionic and electronic structures outside the
CuO$_2$ planes have important effects on the low-energy electronic
states and the superconducting properties \cite{Attfield_Nature98,
Eisaki, McElroy, RichardP}. In particular, the substitution of
Sr$^{2+}$ by Ln$^{3+}$ (Ln = La, Pr, Nd, Sm, Eu, Gd, Bi) in
Bi$_2$Sr$_{2-x}$Ln$_x$CuO$_{6+\delta}$ leads to a critical
temperature that significantly depends on ion radius mismatch
($\Delta$$r$): at $x = 0.4$, $T_c$ $\sim$ 30K for La-substitution
which has the smallest $\Delta$$r$, while $T_c$ $\sim$ 0K for
Bi-substitution which has the largest $\Delta$$r$ \cite{Eisaki}.
These results have been cited as evidence of strong dependence of
$T_c$ to apical site (A-site) disorder. However, our results suggest
that this is not the complete story. As shown in
Fig.~\ref{mapping}h, at the fixed substitution level $x = 0.4$, the
hole doping levels are different for La-Bi2201 ($p \sim 0.16$,
optimally doped) and Bi-Bi2201 ($p \sim 0.05$, heavily underdoped).
Nevertheless, there is a factor of 3 in terms of the maximum $T_c$'s
in these two systems, suggesting the superconducting properties of
Bi$_2$Sr$_{2-x}$Ln$_x$CuO$_{6+\delta}$ are influenced by both charge
underdoping and lattice disorder, the dual characters associated
with chemical substitution.

%Fig 2 ########################################################################
\begin{figure}[htbp]
\begin{center}
\includegraphics[width=8cm]{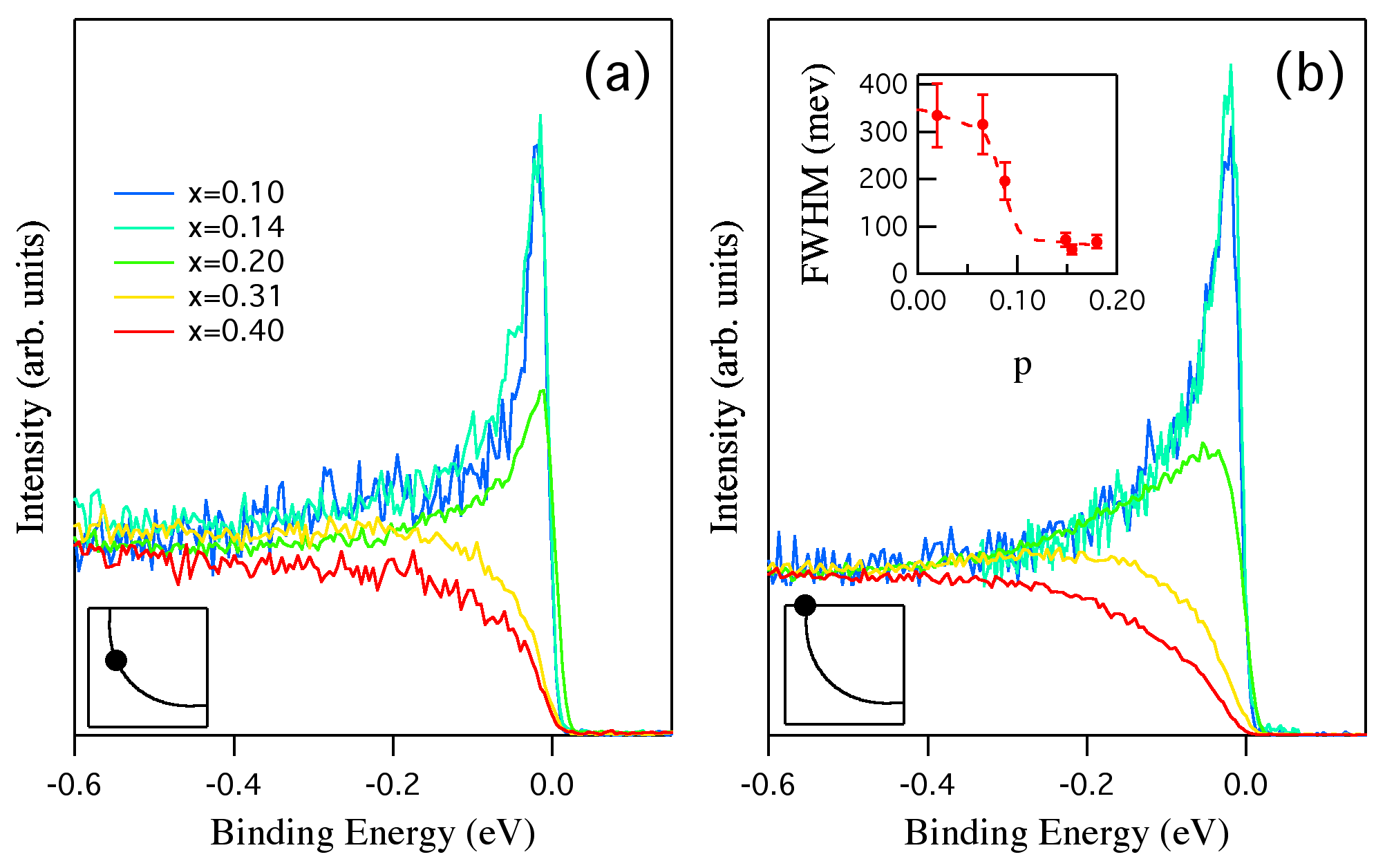}
\caption{\label{shape} Comparison of EDC lineshape of Bi-Bi2201 at different Bi
contents. (a) Near the node, and (b) near the antinode. The inset in
(a)  and (b) with black color show the k positions where those EDCs
are taken. The inset in red color in (b) shows
full-width-at-half-maximum (FWHM) of antinodal EDCs at K$_F$. The
dashed line is a guide to eyes.}
\end{center}
\end{figure}
% #######################################################################

Another interesting phenomenon we observed is the dependence of the
quasiparticle (QP) spectral coherence  on the Bi content. In
Fig.~\ref{shape}, we compare the near-nodal and antinodal energy
distribution curve (EDC) lineshape for various Bi contents. Sharp
EDC peaks are observed in both directions at low Bi substitution
levels. The peak broadens as $x$ increases and is not observed in
the heavily substituted samples. This loss of coherence is also
accompanied by an opening of a soft energy gap characterized by the
suppression of spectral weight in the vicinity of the Fermi energy
(E$_F$).  The crossover between the two regimes occurs around $x$ =
0.20, which corresponds to a doping of $p$ $\sim$ 0.1, the
superconducting-nonsuperconducting phase boundary at zero
temperature.  We plot the width of EDCs at k$_F$ for the antinodal
region in the inset of Fig.~\ref{shape}b. One clearly sees a
significant linewidth broadening when $p < 0.1$, corresponding to
$x > 0.2$. This may suggest that the superconductivity is closely
correlated to the QP coherence. We also note that the opening of
this soft gap first appears in the antinodal region and spreads out
to the nodal one upon increasing substitution, as indicated in
Fig.~\ref{shape}. A similar but smaller nodal gap has been also
observed previously in several lightly doped high-$T_c$ cuprates,
and was attributed to disorder effect \cite{KMShen}.

%Fig 3 ######################################################################
\begin{figure}[htbp]
\begin{center}
\includegraphics[width=8cm]{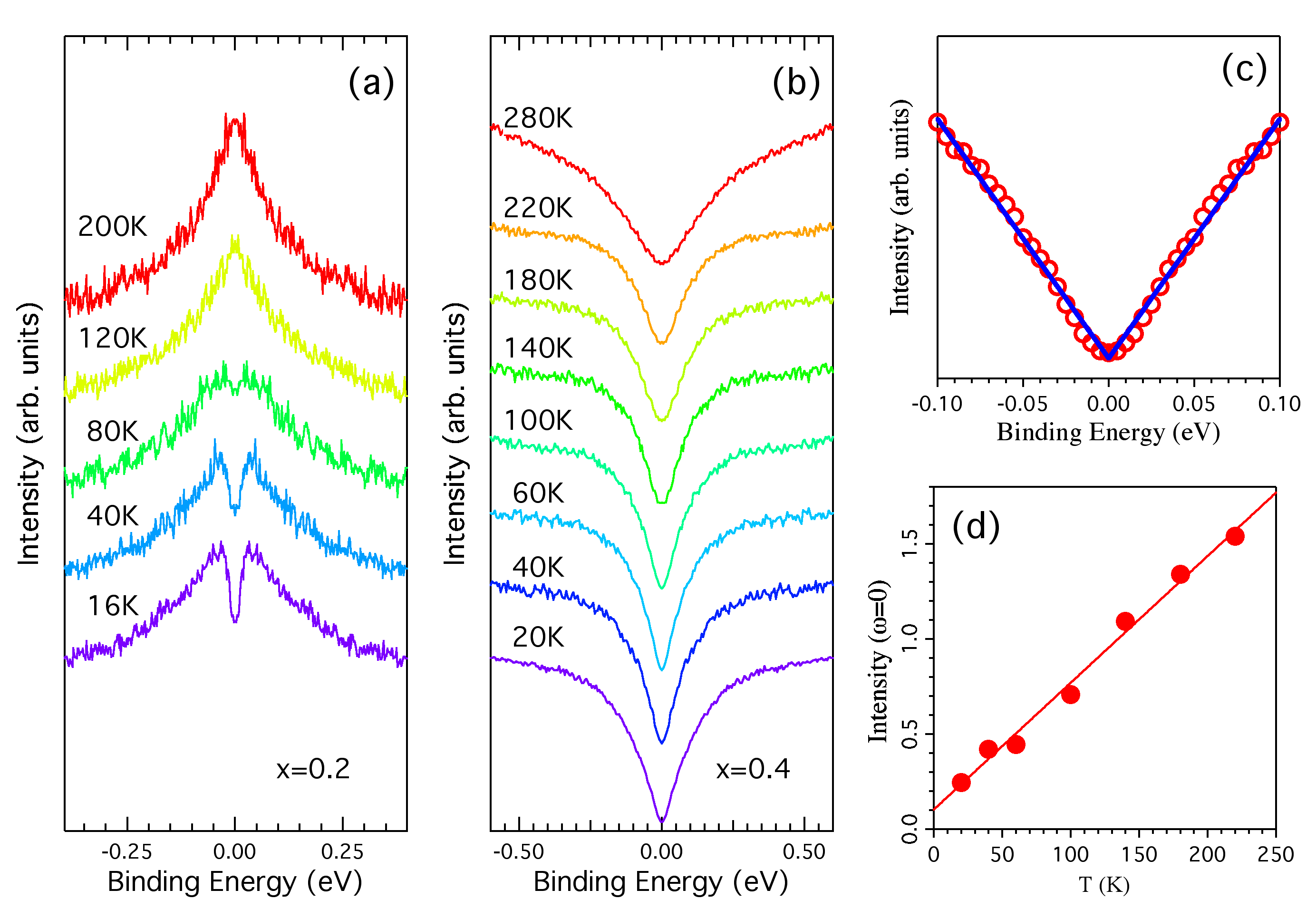}
\caption{\label{twogaps} (a) Temperature dependence of symmetrized EDCs at
K$_F$ in the antinodal region for $x$ = 0.2 sample, and (b) for $x$
= 0.4 sample. (c) Zoom-in view of the 20K EDC shown in (b)
near E$_F$ and a linear fit to it. (d)  Temperature dependence of
EDC intensity at E$_F$ (dots) and the corresponding linear fit
(line) for $x$ = 0.4 sample. }
\end{center}
\end{figure}
% #######################################################################

Perhaps the most surprising finding from our study is the
observation of different antinodal pseudogap behaviors at low and
high Bi contents. The temperature dependence of the symmetrized
antinodal EDCs is given in Figs.~\ref{twogaps}a and b for $x$ = 0.2
and $x$ = 0.4 samples, respectively. The antinodal gap
is filled in and disappears above 80 K for the $x$ = 0.2 sample,
similar to the conventional behavior of the pseudogap observed by
STM for similar samples \cite{STM}. The larger  gap of  the $x =
0.4$ sample behaves differently in the following ways: (1) it lacks
a well-defined spectral structure to define the value of gap,
although the suppression of spectral intensity starts from $\sim$
0.2 eV; (2) while its overall spectral shape is little affected
within the temperature range up to 280 K, the intensity at E$_F$ is
linearly proportional to temperature, as shown in
Fig.~\ref{twogaps}d; (3) the spectral intensity is suppressed to
zero at E$_F$ linearly in energy at low temperature, as shown in
Fig.~\ref{twogaps}c. This reminds us of the classical Coulomb gap (CG)
behavior in strongly disordered two-dimensional systems
\cite{Efros}, and suggest that the pseudogap in the underdoped
regime has evolved into a Coulomb gap following the
superconductor-insulator transition.

\begin{figure}[htbp]
\begin{center}
\includegraphics[width=8cm]{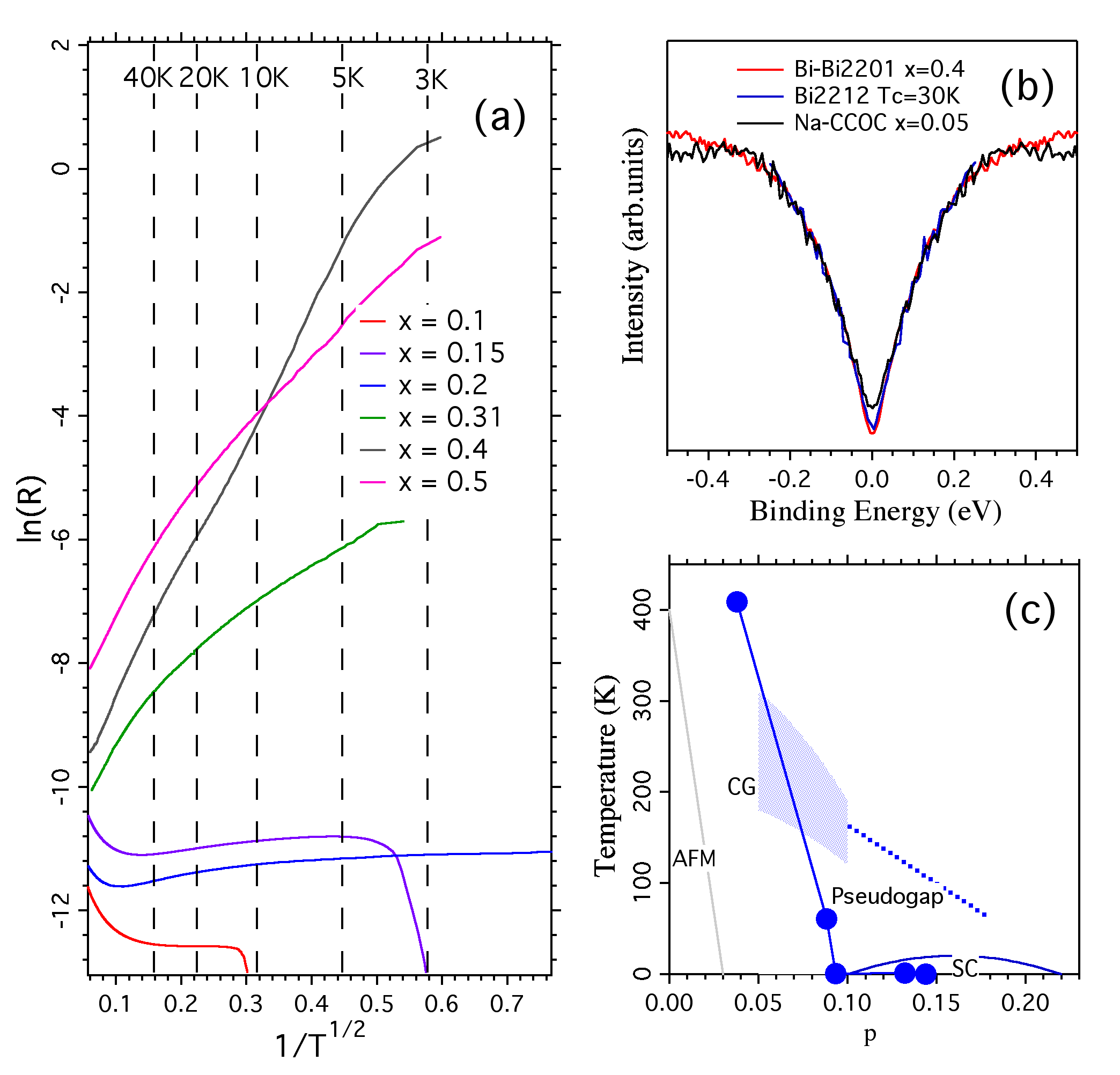}
\caption{\label{phasediagram} (a) Natural logarithm of resistivity versus
$T^{-\frac{1}{2}}$ for samples $x$ = 0.1, 0.15, 0.2, 0.31, 0.4 and
0.5. (b) Lineshape comparison of the antinodal EDCs in
Bi$_{2}$Sr$_{1.6}$Bi$_{0.4}$CuO$_{6+\delta}$,
Bi$_{2.1}$Sr$_{2}$Ca$_{1-x}$Y$_x$Cu$_2$O$_{8}$ (from
Ref.\cite{Tanaka}) and Ca$_{1.95}$Na$_{0.05}$CuO$_2$Cl$_2$ (from
Ref.\cite{shen}). (c) Qualitative phase diagram of
Bi$_{2}$Sr$_{2-x}$Bi$_x$CuO$_{6+\delta}$.}
\end{center}
\end{figure}

To test this finding, we performed transport
measurements on the same samples to look for evidence of the Coulomb
gap. The temperature dependence of the in-plane resistivity of
Bi$_{2}$Sr$_{2-x}$Bi$_x$CuO$_{6+\delta}$ at different Bi content $x$ is
plotted in Fig.~\ref{phasediagram} as the natural logarithm of
resistivity versus $T^{-\frac{1}{2}}$.  The superconductor-insulator
transition around $x=0.2$ is clearly visible. For $x>0.2$, the
resistivity curves show approximately linear behavior over a wider
range of temperature, consistent with the classical hopping
resistivity in the presence of a Coulomb gap,
$\rho(T)=\rho_{0}$exp$(T_0/T)^{\frac{1}{2}}$, expected for a
disordered insulating system with long-range Coulomb interaction
\cite{Efros}. Here $T_0 = e^2/\kappa\xi$ is the long-range Coulomb
energy scale determined by the dielectric constant $\kappa$ and the
localization length $\xi$. From the slope of $T$-dependent curves in
Fig.~\ref{phasediagram}a, we obtain $T_0$ and plot them in
Fig.~\ref{phasediagram}c (red dots). For the $x$=0.4 sample, $T_0
\sim$ 400K, which explains qualitatively the soft gap in the
antinodal spectrum visible even at $T$ = 280K shown in
Fig.~\ref{twogaps}b. Indeed, extrapolating the linear $T$-dependence
of ARPES intensity at E$_F$ (shown in Fig.~\ref{twogaps}d) allows a
rough estimate of the crossover temperature ${T_{CG}^{\ast}} \sim$
500 K above which the suppression of the spectral weight (``gap") is
completely filled. Thus the crossover temperature
(${T_{CG}^{\ast}}$) obtained from ARPES is consistent with $T_0$
derived from the resistivity, suggesting that the same Coulomb
energy scale is involved in both ARPES and transport. 

The ARPES spectra shown in Fig.~\ref{twogaps}b
allows an estimate of the magnitude of the energy gap at the
antinode for the $x$=0.4 sample, $\Delta \sim$ 0.19 eV. This,
combined with the derived value of $T_0 \sim$ 400K for the same
sample (see Fig.~\ref{phasediagram}c), gives a ratio of $\Delta/T_0
\sim 5$. It is known from the theory of the classical Coulomb gap
\cite{Efros}, $\Delta/T_0 = g_0(\xi)^2T_0$ where $g_0$ is
unperturbed density of states. A reasonable value of $g_0 \sim$ 2.1
state/eV cell was provided by band theory for Bi2201
\cite{LDA_Bi2201}. Taking $\xi \sim$ 8 unit cells, we estimate that
the ratio predicted by the Coulomb gap theory to be $\sim$ 5. This
remarkable qualitative agreement between the experimental ratio and
theoretical one strongly supports the Coulomb gap nature of the
observed soft gap in the photoemission spectra. It is interesting to
note that $T_0$ becomes smaller as the Bi content decreases and
becomes zero as $x$ becomes 0.2, the superconductor-insulator
boundary revealed by the combined effects of underdoping and
disorder. We caution here that the conventional Coulomb gap usually
referred to the linear suppression of the density of states (DOS) in
an isotropic system, not to the spectral function measured near the
antinodal region of a highly anisotropic material. Nevertheless, the
antinodal intensity dominates the DOS in the hole-doped cuprates due
to the van Hove singularity. 

%Fig 4 #########################################################################
% #######################################################################

A natural question to ask is how universal this Coulomb gap behavior
is in the heavily underdoped high-$T_c$ cuprates achieved by
chemical substitution. To answer this question, we compared the
antinodal EDC of the $x = 0.4$ sample with heavily underdoped
Bi$_{2.1}$Sr$_{2}$Ca$_{1-x}$Y$_x$Cu$_2$O$_{8}$ ($T_c$ $\sim$ 30K)
\cite{Tanaka} and nonsuperconducting
Ca$_{1.95}$Na$_{0.05}$CuO$_2$Cl$_2$ (Na-CCOC) \cite{shen}, as shown
in Fig.~\ref{phasediagram}b. We found that the lineshape of these
different samples are very similar, characterized by the opening of
a large soft gap and the absence of the QP peak.

Interestingly, as shown in the case of
Bi$_{2.1}$Sr$_{2}$Ca$_{1-x}$Y$_x$Cu$_2$O$_{8}$ \cite{Tanaka}, the
antinodal leading-edge gap was found to increase upon more
underdoping, while the near-nodal gap seems to decrease and be
proportional to $T_c$. This apparent gap dichotomy (or two-gap
scenario) has been interpreted as evidence that the $d$-wave pairing
gap opens along the FS arc around the node while the
antinodal gap is a different type of gap that may not contribute to
superconductivity \cite{Tanaka}. However, with the appearance of a
Coulomb gap in heavily substituted/underdoped samples, we raise
another possibility, as summarized in the schematic phase diagram
shown in Fig.~\ref{phasediagram}c: upon increasing underdoping, the
$d$-wave-like leading-edge gap region may be influenced or even
truncated by the Coulomb gap region, which by itself is induced by
the intrinsic disorder associated with chemical substitution. While
this Coulomb gap region might push the onset of the superconducting
phase to a higher doping level, as observed in the substituted
Bi2201, it may also result in a spin glass phase often observed
between the antiferromagnetic insulating phase and the
superconducting state at low temperatures.

This work was supported
by grants from the US NSF DMR-0353108, DMR-0704545, and DOE
DEFG02-99ER45747. This work is based upon research conducted at the
Synchrotron Radiation Center supported by NSF DMR-0537588, and the
Advanced Light Source supported by DOE No. DE-AC02-05CH11231. The
work at the IOP, Beijing was supported by the NSFC, the MOST 973
project (No. 2006CB601000, 2006CB921802), and the CAS project
ITSNEM.

\bibliography{biblio_en}

\end{document}